\documentclass[aps,twocolumn,nofootinbib]{revtex4-1}
\usepackage{graphicx}
\usepackage{hyperref}
\usepackage{amsmath}
\usepackage{amssymb}

\newcommand{\beq}{\begin{eqnarray}}
\newcommand{\eeq}{\end{eqnarray}}
\def\simlt{\stackrel{<}{{}_\sim}}

\begin{document}

\title{Dark Matter as a Trigger for Periodic Comet Impacts}
\author{{Lisa Randall and Matthew Reece}}
\affiliation{Department of Physics, Harvard University, Cambridge, MA, 02138}

\begin{abstract}
Although statistical evidence is not overwhelming, possible support for an  approximately 35 million year periodicity in the crater record on Earth could indicate a nonrandom underlying enhancement of meteorite impacts at regular intervals. A proposed explanation in terms of tidal effects on Oort cloud comet perturbations as the Solar System passes through the galactic midplane  is hampered by lack of an underlying cause for sufficiently enhanced gravitational effects over a sufficiently short time interval and  by the time frame between such possible enhancements. We show that a smooth dark disk  in the galactic midplane would address both these issues and create a periodic enhancement of the sort that has potentially been observed.  Such a disk is motivated by a novel dark matter component with dissipative cooling that we considered in earlier work. We show how to evaluate the statistical evidence for periodicity  by input of appropriate measured priors from the galactic model, justifying or ruling out periodic cratering with more confidence than by evaluating the data without an underlying model. We find that, marginalizing over astrophysical uncertainties, the likelihood ratio for such a model relative to one with a constant cratering rate is 3.0, which moderately favors the dark disk model.  Our analysis furthermore yields a posterior distribution that, based on current crater data, singles out a dark matter disk surface density of approximately 10 $M_\odot/{\rm pc}^2$. The geological record thereby motivates a particular model of dark matter that will be probed in the near future.
\end{abstract}

\maketitle

Large meteorite strikes on Earth cause big impact craters that are very likely responsible for some mass extinctions \cite{Alvarez1980}. Possible evidence of $\approx 35$ million year periodicity in the dates of these events suggest a nonrandom underlying origin \cite{RaupSepkoski84,RampinoStothers84,AlvarezMuller84,WhitmireJackson84,DavisHutMuller84,Hut1987,SchwartzJames84,Stothers98,Yabushita2004,Napier2006,BailerJones:2011zh}. Although not yet clearly established, it is of interest to explore possible underlying causes, especially if they have other measurable consequences. Two suggestions were made simultaneously by multiple groups  to explain a periodic enhancement of Oort cloud induced comets hitting the Earth. One, known as the ``Nemesis hypothesis,'' was that the Sun has a so-far undetected companion star \cite{WhitmireJackson84, DavisHutMuller84}.  No companion has been detected. The other suggestion involves the Sun moving through the plane of the galaxy. The Milky Way, like other spiral galaxies, has a large fraction of its normal (baryonic) matter arranged  in the shape of a flattened disk, with the density falling off exponentially over a characteristic distance of 3 kpc in the radial direction but in a much shorter characteristic distance of about 300 parsecs in the vertical direction \cite{BinneyTremaine,MvdBW}. The flattened shape arises because normal matter cools by emitting photons that carry kinetic energy away from the galaxy. This lowers the velocity of ordinary matter and the less energetic particles move in a smaller volume due to their reduced velocities and their gravitational interactions. Such particles do however retain angular momentum, so the phase space doesn't shrink in the radial direction. Matter therefore forms a flattened disk with small vertical height.  

The idea for explaining periodic cratering  is that the Sun, as it orbits the Galactic Center, oscillates up and down through the plane of the galaxy, leading to periodic perturbations of the Oort cloud from enhanced density near the plane. These perturbations cause comets to enter the inner Solar System resulting in comet showers \cite{RampinoStothers84,SchwartzJames84}. However, to date no suggested mechanism for the enhanced density is successful in explaining the timing and magnitude of the periodicity.  Molecular clouds have been suggested \cite{RampinoStothers84,SchwartzJames84} but they have been shown to be spread too far from the plane to justify periodic cratering \cite{ThaddeusChanan85}. The period is in any case too short to be accounted for by conventional baryonic matter, which as mentioned above also does not have a large enough vertical density gradient to explain a strong periodic signal. Remarkably, a dark matter disk could address both these issues. 

Despite the apparent lack of fundamental explanation, studies have searched for periodic phenomena by fitting {\em ad hoc} sinusoidal templates without an underlying physical model. These were recently reviewed in Ref.~\cite{BailerJones:2011zh}. Recent analyses of the crater data usually find that a period of about 35 Myr is most consistent with the data, although the statistical evidence is weak and disappears completely when the look elsewhere effect is taken into account (if there is no prior favoring particular periods). In this letter,  we conjecture that thin dark matter disks, which would form if a species of dark matter has dissipative dynamics~\cite{Fan:2013yva}, could affect meteorite impacts and address both of the above issues.
 The bulk of dark matter, based on observed rotation curves and expected properties of weakly interacting particles, is known to be arranged in a roughly spherical halo, gradually growing less dense over distances of order 20 kpc. However, this has been established only for the majority of dark matter. A small fraction might have interactions similar to those of baryons, emitting ``dark photons'' and dissipating energy, thereby cooling into an even thinner dark disk embedded in the ordinary baryonic disk \cite{Fan:2013yva}.  The existence or nonexistence of such a disk will be probed most directly over the next decade through extensive measurements of stellar kinematics in the Milky Way \cite{Famaey:2012ga,Rix:2013bi}.  Assuming the dominant perturbing mechanism is the tidal force, which is proportional to the density of the disk \cite{HeislerTremaine1986}, the Sun's passage through the dark matter disk would  cause enhanced periodic Oort cloud perturbations. We find that the observed crater dates agree with such a model better than with a constant cratering rate by a likelihood ratio of 3.0, and single out a dark matter disk surface density of approximately 10 $M_\odot/{\rm pc}^2$.  This proposal will be tested by upcoming measurements from the Gaia satellite that will narrow the range of priors, and hence the possible cratering predictions. More precise measurements of the Milky Way's properties will thereby provide a sharper statistical test of the comet shower hypothesis.  The results could ultimately reveal  a strong  dark matter influence on the history of our Solar System and even of life here on Earth.

We reframe the problem of testing galactic influences on the terrestrial impact crater record in a form that is more robust than testing the data for periodicity. The observation of possible periodicity was an important impetus for the original hypotheses of astrophysical influences on life on Earth. However,  the science will be vastly improved by setting priors with current and future data about our galaxy. We show how to use all available measured data to pin down the shape of the galaxy and derive a detailed trajectory of the Sun as it oscillates through.

\begin{figure}[!h]\begin{center}
\includegraphics[width=0.85\columnwidth]{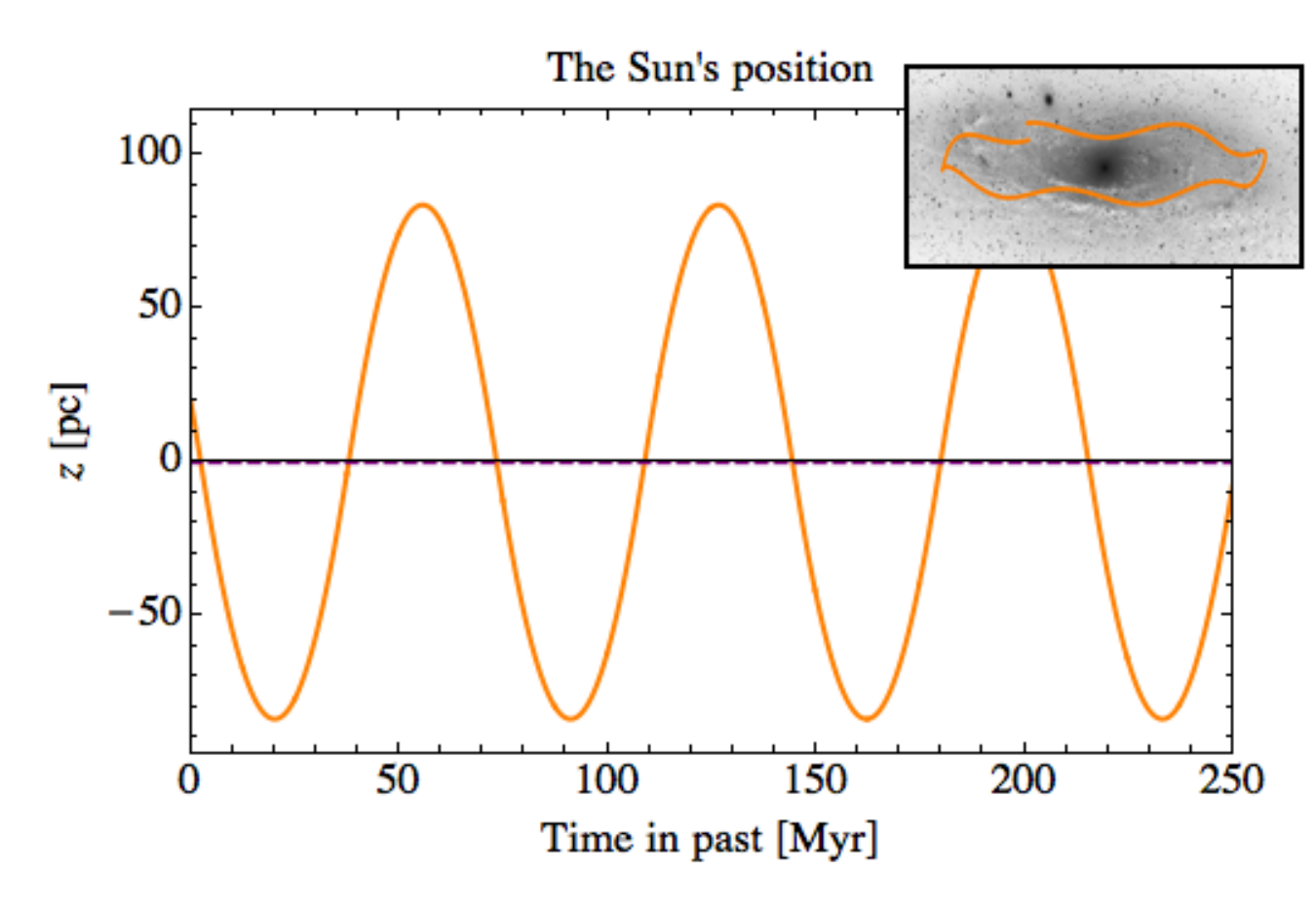}
\end{center}
\caption{The Sun's height above the galactic plane as a function of time, extrapolated backward via Eq.~\ref{eq:zaccel}. The corresponding cratering probability is shown in Fig.~\ref{fig:goodfit}. Inset: an illustration of how the Sun moves around the galactic center while also oscillating vertically; the vertical oscillation is exaggerated for visibility.}\label{fig:sunheight}
\end{figure}%

Our focus in this letter is the influence of a dark disk in this context. We take the mass distribution of the disk to be an isothermal sheet, with density:
\beq
\rho_{\rm disk}(R,z) = \frac{M_{\rm disk}}{8\pi R_d^2 z_d} \exp\left(-R/R_d\right) {\rm sech}^2\left(z/\left(2 z_d\right)\right). \label{eq:isothermalsheet}
\eeq 
This form, with density falling exponentially with radius and height, can be derived from the Poisson equation for a gravitationally interacting set of particles that have a Maxwellian vertical velocity distribution~\cite{MvdBW}. In reality, a thin disk will fragment into smaller clouds or clumps, the final size of which depends on details of cooling and angular momentum transfer. We assume their distribution to be uniform enough to approximate a smooth tidal field. We characterize the matter in a disk via its {\em surface density} $\Sigma$, which is the integral of $\rho(R,z)$ over $z$ at fixed radius $R$. We assume an equal scale radius for baryons and dark disk matter, $R_d \approx 3$ kpc \cite{BinneyTremaine}. We use a one-dimensional model of the Sun's motion through the galaxy, assuming small vertical oscillations around a circular orbit, with acceleration determined by the local density:
\beq
a(z) \equiv \ddot{z} = - \frac{\partial \Phi}{\partial z} \approx - \int dz~4\pi G\rho(z) \label{eq:zaccel}.
\eeq
This equation relies on the fact that the Milky Way's rotation curve is flat at the radius of the Sun's orbit, so that $\frac{\partial}{\partial R}\left(R \frac{\partial\Phi}{\partial R}\right) \approx 0$. An example of the vertical motion derived in this approximation is shown in Fig.~\ref{fig:sunheight}.

\begin{figure}[!h]\begin{center}
\includegraphics[width=0.65\columnwidth]{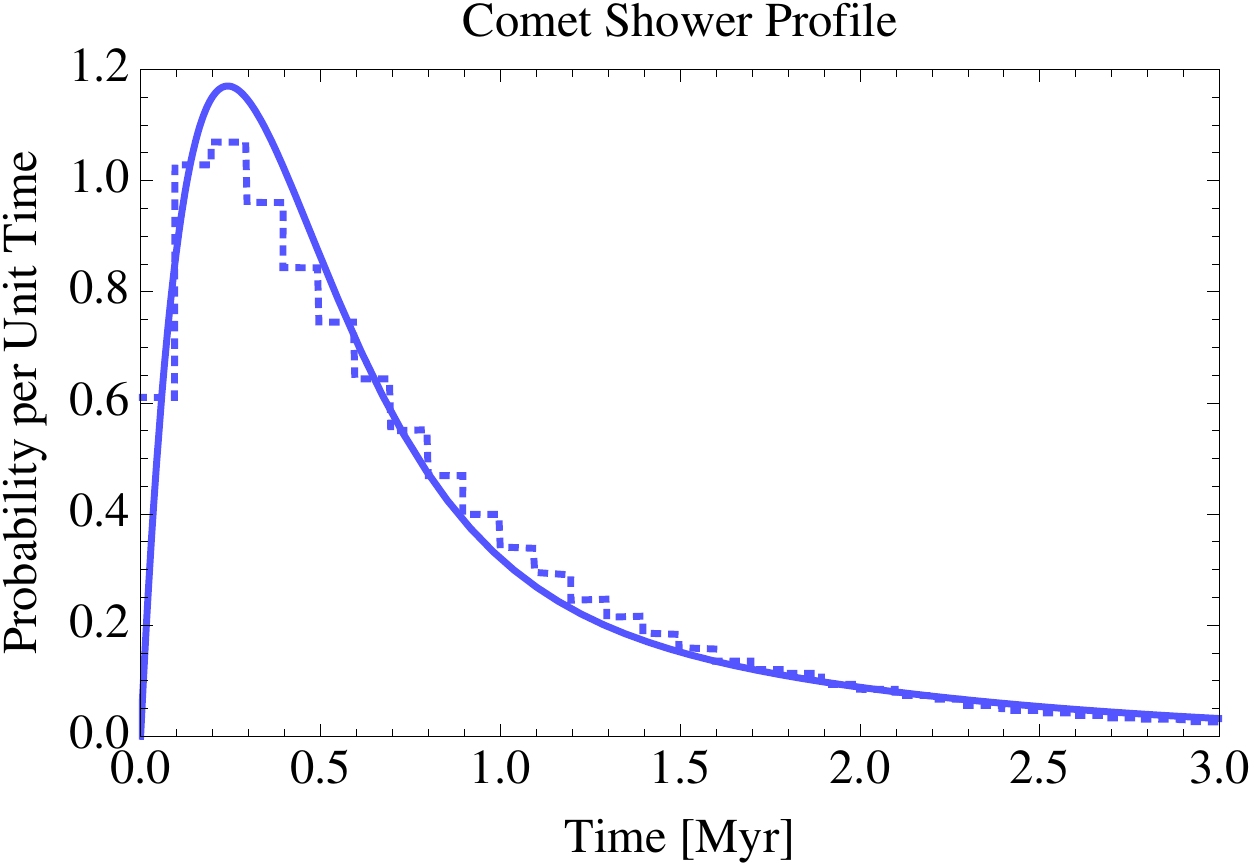}
\end{center}
\caption{Comet shower profile: dashed line from~\cite{Hut1987}; solid line an ansatz we use in the numerics.}\label{fig:cometshowerprofile}
\end{figure}

We assume, as a first approximation, that the probability that a comet shower begins at a time $t$ is proportional to the total local matter density near the Sun at that time, $\rho(t)$. This assumption is motivated by Refs. \cite{HeislerTremaine1986,HeislerTremaineAlcock1987,Fernandez2002}, which argue that perturbations to the Oort cloud are a result of tidal forces. The initial paper by Heisler and Tremaine~\cite{HeislerTremaine1986} demonstrates that tidal effects dominate over stellar perturbations. However, because they assume a uniform disk, comet showers came only from the combined effects of stars and the tide~\cite{HeislerTremaineAlcock1987} and occur only infrequently. With a dark disk the tidal effect still dominates, and with a thin disk the temporal variation can suffice to explain even a 35 million year period. The tidal forces gradually alter the angular momentum of the comet by modifying its transverse velocity $v_T$: up to factors depending on time-dependent angles, $dv_T/dt \sim r \partial_z^2 \Phi(z) \sim 4\pi G r\rho$, with $\Phi$ the gravitational potential and $r$ the Sun-comet distance. Comets with $v_T$ small enough compared to the circular velocity move on approximately radial orbits, falling into the inner Solar System. Thus, tidal forces gradually strip comets with small transverse velocities out of the Oort cloud at a rate proportional to the local density at any given time. These comets near the edge of the loss cone enter the inner Solar System in a time of order their orbital time of $\simlt 1$ Myr, which is less than or approximately the time of transit of the dark disk. We model this time delay based on a published result that used Monte Carlo simulation to deduce the longevity of the perturbation's influence~\cite{Hut1987}, illustrated in Fig.~\ref{fig:cometshowerprofile}. The convolution of this time delay with the density $\rho(t)$ near the Sun defines $r(t)$, the rate for impact craters at time $t$.

We confront the model with observations of craters listed in the Earth Impact Database \cite{EarthImpactDatabase}. We (arbitrarily) choose to focus on craters greater than 20 km in diameter (since smaller craters occur much more frequently and don't necessarily require large comet-induced impacts) within the last 250 million years (as  a minimal way to model the fact that older craters are eroded and rarely found). Ultimately we would want to be able to distinguish impacts due to asteroids vs.~comets to obtain a better test of the hypothesis. There is also data on $^{3}$He in dust from comets that can ultimately lend support to (or refute) an assumed periodicity \cite{Farley1998}. We find this possibility exciting but neglect this data for the time being.

We focus on the {\em likelihood ratio} between two models,
\beq
\frac{P\left({\rm data}|{\rm model}_1\right)}{P\left({\rm data}|{\rm model}_2\right)},
\eeq
which via Bayes' theorem is also the ratio of posterior probabilities for the models if we begin with equal prior probabilities. We are interested in whether the data provide support to a model in which the rate of impact crater events over time, $r(t)$, is driven by the Sun's motion through the galaxy. We compute the likelihood $P({\rm data}|{\rm model})$ as a product of the probabilities for each event, which are given by the overlap of the Gaussian characterizing the observed crater age with the model's rate function $r(t)$:
\beq
P_{\rm event}(E_i) = \int_{t_{\rm begin}}^{t_{\rm end}}\frac{ (t_{\rm end} - t_{\rm begin})r(t)}{\sqrt{2\pi}\sigma_i}e^{-\frac{(t - t_i)^2}{2\sigma_i^2}} dt. \label{eq:probevent}
\eeq 
The factor $t_{\rm end} - t_{\rm begin}$ is present so the result will be dimensionless; we will compare ratios of likelihoods, so this factor will drop out. For any given set of galactic parameters, we normalize $r(t)$ so that the average expected number of craters in 250 Myr matches the number in our sample, which provides an optimal fit. Ideally, in the future a detailed model of the Oort cloud would specify the normalization of $r(t)$, in which case a factor $P_{\rm gap}(t_0, t_1) = \exp\left(-\int_{t_0}^{t_1} r(t) dt\right)$ is required.

The likelihood ratio allows us to quantify the evidence for a hypothesis relative to a different hypothesis, which we take to be a constant probability per unit time.  The periodic fits in the literature to date with a period of about 35 million years match the data better than an assumed constant rate of meteorite hits, but the statistical significance seems to disappear when the ``look elsewhere'' effect is taken into account \cite{BailerJones:2011zh}. That is, there are so many possible periodic functions that the fact that some do better is not a significant result. That conclusion changes when a model with measured priors is used, rather than a random periodic model. Constraints on the galactic density select a range of reasonable periods.

\begin{table}
\begin{center}
\begin{tabular}{lll}
Parameter & Prior\\
\hline
$\Sigma_B^{1.1} \equiv \Sigma_B(\left|z\right| < 1.1~{\rm kpc})$ & Gaussian, $55 \pm 5~M_\odot/{\rm pc}^2$~\cite{Zhang:2012rsb}\\
$z_d^B$ & Gaussian, $300 \pm 60~{\rm pc}$~\cite{Juric:2005zr}\\
$\Sigma_D$ & Flat from 0 to $30~M_\odot/{\rm pc}^2$\\
$z_d^D$ & Flat on $\log z_d^D$ from 0.1 pc to 1 kpc \\
$\rho_{\rm halo}$ & Gaussian, $0.3 \pm 0.1$ GeV/cm$^3$~\cite{Bovy:2012tw}\\
$Z_\odot$ & Gaussian, $26 \pm 3~{\rm pc}$~\cite{Majaess:2009xc}\\
$W_\odot$ & Gaussian, $4.04 \pm 0.48~{\rm pc/Myr}^2$~\cite{Schoenrich:2009bx}\\
$\Sigma_{\rm tot}^{1.0} \equiv \Sigma_{\rm tot}(\left|z\right| < 1.0~{\rm kpc})$ & Gaussian, $67 \pm 6~M_\odot/{\rm pc}^2$~\cite{Zhang:2012rsb}
\end{tabular}
\end{center}
\caption{Summary of the factors making up the prior probability distribution.}\label{tab:prior}
\end{table}

\begin{figure}[!h]\begin{center}
\includegraphics[width=1.0\columnwidth]{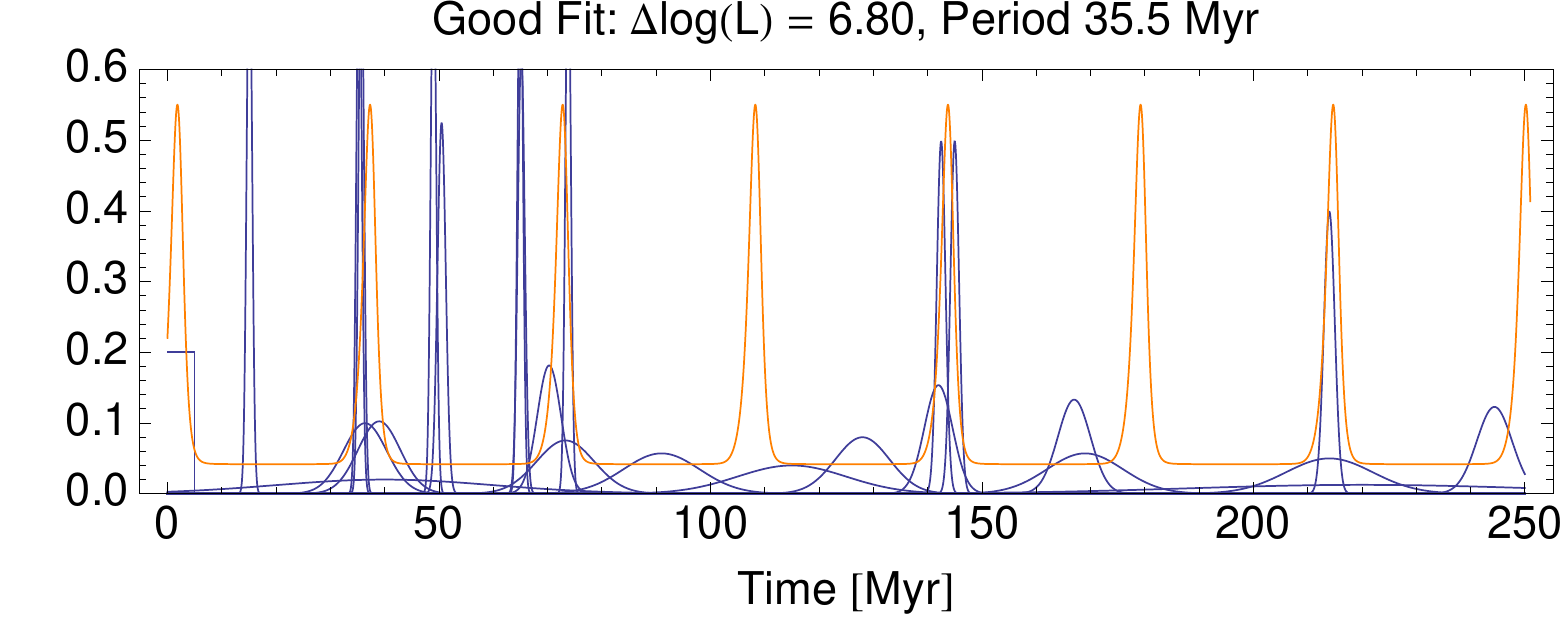}
\end{center}
\caption{An example of a model that provides a good fit. The parameters of the dark disk are $\Sigma_D = 13 M_\odot/{\rm pc}^2$ and $z_d^D = 5.4~{\rm pc}$. The baryonic disk is 350 pc thick with total surface density 58 $M_\odot/{\rm pc}^2$. The local dark halo density is 0.037 GeV/cm$^3$. 
$Z_\odot = 20$ pc and $W_\odot$ = 7.8 km/s. In this case, the period between disk crossings is about 35 Myr. In orange is the rate $r(t)$ of comet impacts (with arbitrary normalization). This is approximately proportional to the local density, but convolved with the shower profile from Fig.~\ref{fig:cometshowerprofile}. The various blue curves each correspond to one recorded crater impact.}\label{fig:goodfit}
\end{figure}%

\begin{figure*}[t]\begin{center}
\includegraphics[width=0.8\textwidth]{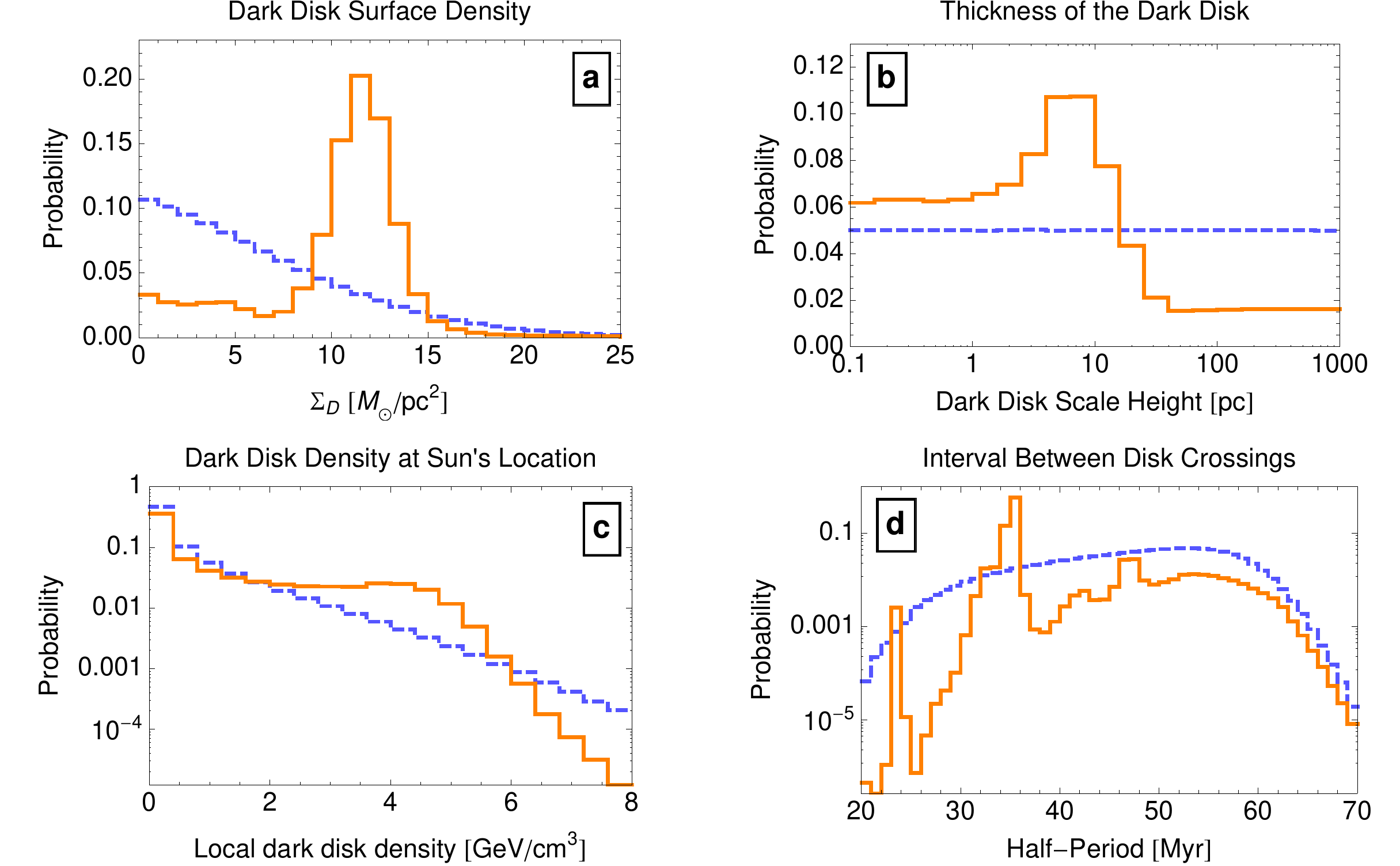}
\end{center}
\caption{Preferred parameters. One-dimensional projections of the prior (blue, dashed) and posterior (orange, solid) probability distributions. (a) The surface density of the dark disk, which the posterior distribution prefers to be between about 10 and 15 $M_\odot/{\rm pc}^2$. (b) The dark disk thickness, which fits best at about 10 parsec scale height but extends to thinner disks. The posterior distribution is flat even for very thin disks, because comet showers last for around a million years even if the Solar System passes through the disk in a shorter time. (c) The local density of disk dark matter (relevant for solar capture or direct detection), which has significant weight up to several GeV/cm$^3$. (d) The interval between times when the Sun passes through the dark disk, which fits best at values of about 35 Myr.}\label{fig:probdist}
\end{figure*}%

Our assumed parameters are:  baryonic disk parameters $\Sigma_B$ and $z^B_d$, the dark disk parameters $\Sigma_D$ and $z^D_d$, the dark halo parameter $\rho_{\rm halo}$, the Sun's position $Z_\odot$ and velocity $W_\odot$.  Collectively, these seven quantities parameterize the model, and to assign a likelihood we marginalize over them, i.e. integrate over the space of parameters weighted by the prior distribution. The seven parameters are straightforwardly related to the first seven constrained quantities in Table 1, with one extra constraint on the total surface density. (By Eq.~\ref{eq:isothermalsheet}, $\Sigma_B = \Sigma_B^{1.1}/\tanh(1.1~{\rm kpc}/(2 z_d^B))$.) We sample random numbers directly from the distributions in each row of Table 1 except for $\Sigma_{\rm tot}^{1.0}$. We then compute this total density for the sample parameters and apply Monte Carlo sampling (keeping the point if a random number is less than the weight assigned to $\Sigma_{\rm tot}^{1.0}$ in the last line of the table). Thus, in the end $\Sigma_D$ does not have a flat distribution, but has been reweighted to penalize choices with too much total density.

After marginalizing over all parameters, we find a likelihood ratio of 3.0 for the dark disk model compared to a uniform cratering rate. In other words, if we assigned equal prior probabilities, then in light of the data our model is more likely by a factor of 3. This Bayes factor is not large enough to be decisive, but it is intriguing. It indicates we should find the dark disk moderately more plausible than we did a priori. An example of parameters with larger likelihood is shown in Fig.~\ref{fig:goodfit}. 

Although the likelihood ratio favors the model of a dark disk over a uniform rate, it does not tell us if either fits well. Hence, we perform a Cram\'er--von Mises test to find a $p$-value for the data (comparing empirical and theoretical cdfs). For constant $r(t)$, we find $p \approx 0.09$. For the model in Fig.~\ref{fig:goodfit}, this improves to $p \approx 0.13$. Thus, these models give reasonable (but not perfect) fits to the data: we cannot reject them at 95\% confidence level. As such, it makes sense to compare them, and the likelihood ratio gives a mild preference to the disk model. For a different perspective we consider the Akaike information criterion~\cite{AkaikeInformationCriterion} as modified for small sample sizes~\cite{HurvichTsai}. This compares maximum $\log{\cal L}$ but  penalizes models with more parameters: in our case seven parameters with a dark disk versus five without. Our maximum likelihood difference is $\Delta \log {\cal L} \approx 6$ when $\Sigma_D \approx 13~M_\odot/{\rm pc}^2$, and the modified AIC criterion asks for $\Delta \log {\cal L} > 3.6$, so again we find a preference for the dark disk model.

Furthermore, a Bayesian analysis makes predictions for the values of parameters that can be measured in the future. We show the prior and posterior distributions for a few of our parameters in Figure~\ref{fig:probdist}. The posterior distribution strongly favors a dark disk surface density of $\Sigma_D \sim 10~M_\odot/{\rm pc}^2$ and scale height $z_d^D \sim 10$ pc. These parameters are not yet tested, but involve a large enough dark matter disk density that we expect measurements of stellar kinematics from the Gaia satellite \cite{Famaey:2012ga,Rix:2013bi} to be a stringent test of the proposal in the near future. Once such measurements are in hand, we can turn the question around and predict a cratering rate, strengthening the link between galactic and terrestrial data.

This dark disk surface density is consistent with current observational constraints once the overall uncertainty in the dynamically determined surface density and the large uncertainty in the ISM is accounted for. The ISM value of 13~$M_\odot/{\rm pc}^2$~\cite{Bovy:2013raa} includes 2~$M_\odot/{\rm pc}^2$ of hot gas and furthermore has an uncertainty that is not precisely given but can be reasonably taken as 50\%~\cite{Holmberg:1998xu}.\footnote{We thank Jo Bovy and Ruth Murray-Clay for discussions on this point.} Furthermore more recent textbooks~\cite{BinneyMerrifield} and~\cite{Tielens} give values of 5.5 and 7.6~$M_\odot/{\rm pc}^2$ respectively.  The argument against a dark disk in~\cite{Bovy:2013raa} did not include this source of uncertainty.\footnote{Again we thank Jo Bovy for discussions.}

The posterior distribution for the current volume density of dissipative dark matter near the Sun peaks at low values but is significant and relatively flat between 1 and 5 GeV/cm$^3$. These densities are significantly larger than those generally assumed in direct detection experiments on the basis of a spherical dark matter halo, leading to interesting model-dependent prospects for detecting low-energy nuclear recoils induced by disk dark matter \cite{Bruch:2008rx,McCullough:2013jma,DDDMdirect}.
 
We will present details elsewhere of a study with disks not necessarily aligned, although we find the new parameters do not lead to a larger likelihood ratio.

We conclude that if a dark disk exists, it could play a significant role in explaining the observed pattern of craters, and possibly even mass extinctions. We have also demonstrated how to use measurements of the galaxy and Solar System to weight models with different parameters and ascertain the statistical significance of our hypothesis. With the prospect of better data that will further constrain the model in the future, the statistical tests will become even more stringent, validating or ruling out our proposal. Meanwhile we find this a fascinating possibility worthy of further exploration. Even though crater data is hard to come by, data about the galaxy will be much more abundant in the near future. When we pin this down we will be better able to unambiguously predict the motion of the Solar System and thereby constrain possibilities for nonrandom structure in crater timing.

{\bf Acknowledgments.} We thank Paul Davies for suggesting this intriguing idea, and for subsequent related correspondence. We thank Oded Aharonson, Jo Bovy, Matthew Buckley, Sean Carroll, Ken Farley, Marat Freytsis, Fiona Harrison, Arthur Kosowsky, Eric Kramer, David Krohn, Avi Loeb, Ruth Murray-Clay and Scott Tremaine for useful discussions or correspondence. We thank the referees for useful comments and suggested references. The work of LR was supported in part by NSF grants PHY-0855591 and PHY-1216270. MR thanks the KITP in Santa Barbara for its hospitality while a portion of this work was completed. The KITP is supported in part by the National Science Foundation under Grant No. NSF PHY11-25915.

\end{document}